# ON THE NATURE OF LOW LUMINOSITY NARROW LINE AGN

Ari Laor
Physics Department, Technion, Haifa 32000, Israel
laor@physics.technion.ac.il

## ABSTRACT

There is clear observational evidence that some narrow line (type 2) AGN have a hidden broad line region (BLR), and are thus intrinsically broad line (type 1) AGN. Does this AGN unification applies for all type 2 AGN? Indirect arguments suggest that some "true" type 2 AGN, i.e. AGN having no obscured BLR do exist, but it is not clear why the BLR is missing in these AGN. Here we point out a possible natural explanation. The observed radius-luminosity relation for the BLR implies an increasing line width with decreasing luminosity for a given black hole mass ($M_{\rm BH}$). In addition, there appears to be an upper limit to the observed width of broad emission lines in AGN of $\Delta v_{\rm max} \sim 25,000$ km s$^{-1}$, which may reflect a physical limit above which the BLR may not be able to survive. Thus, at a low enough luminosity the BLR radius shrinks below the $\Delta v_{\rm max}$ radius, leaving no region where the BLR can exist, although the AGN may remain otherwise 'normal'. The implied minimum bolometric luminosity required to sustain a BLR with $\Delta v < 25,000$ km s$^{-1}$ is $L_{\rm min} \sim 10^{41.8}(M_{\rm BH}/10^8 M_\odot)^2$. All AGN with $L < L_{\rm min}$ are expected to be 'true' type 2 AGN, i.e. narrow line AGN without a hidden BLR. Predictions for the true nature of low luminosity AGN in two samples of nearby galaxies are provided. These can be used to test the above $L_{\rm min}$ conjecture, and the predictions of other models for the size and origin of the BLR.

*Subject headings:* galaxies: nuclei-quasars: general

## 1. INTRODUCTION

Active Galactic Nuclei (AGN) are compact sources of non-stellar continuum and/or non-stellar induced line emission at the centers of galaxies. Apart from BL Lac objects, all other types of AGN display emission lines. Type 1 AGN show broad (FWHM $\gtrsim 10^3$ km s$^{-1}$) permitted lines and narrow (FWHM $\lesssim 10^3$ km s$^{-1}$) forbidden lines, while type 2 AGN show only narrow forbidden lines (e.g. Peterson 1997). It is now well established that at least some type 2 AGN contain hidden type 1 AGN, as indicated by optical spectropolarimetry and near IR spectroscopy (e.g. Antonucci & Miller 1985; Tran 1995; Barth, Filippenko, & Moran 1999; Moran et al. 2000; Veilleux, Sanders, & Kim 1997, 1999). This led to the so called "AGN orientation unification" scheme (e.g. Antonucci 1993). *However, does the "AGN orientation unification" apply for all type 2 AGN?*

Some type 2 AGN show no broad lines in either polarized light or near IR spectroscopy. This may either imply they are "true" type 2 AGN, or it may result from a large column of obscuring material towards the nucleus together with the lack of a properly placed "mirror" (i.e. an unobscured scattering medium). The presence of an obscuring material along our line of sight to the nucleus of type 2 AGN is revealed through X-ray spectroscopy (e.g. Maiolino et al. 1998; Risaliti, Maiolino, & Salvati 1999; Matt et al. 2000; Moran et al. 2001). However, it is not clear whether this absorbing medium necessarily also obscures the BLR, as some type 1 AGN show strong X-ray absorption, and much weaker absorption of the BLR (this may be true for all broad absorption line quasars, e.g. Gallagher et al. 2001; Green et al. 2001). A more general test for the presence of an obscuring material close to the nucleus, but not necessarily along the line of sight, can be obtained by looking for the "waste heat" which must be emitted by the obscuring material, if it resides on scales which can obscure the BLR (Whysong & Antonucci 2003). These tests, together with statistical arguments based on isotropic properties of type 1 vs. type 2 AGN, and on their relative number at low redshift (for radio loud AGN), suggest there are true type 2 AGN (Tran 2001, 2003; Singal 1993; cf. Gu & Huang 2002). *Why do some AGN apparently have no broad emission line region? Do these AGN have a fundamentally different inner structure from "normal" AGN?*

In this paper we point out a possible simple argument for the absence of a BLR in some AGN. This argument is based on some of the observed properties of the BLR, and implies the BLR should cease to exist below a critical luminosity, which depends on the mass of the central massive black hole. The structure of the paper is as follows, in §2 we briefly review the dependence of the BLR size and velocity dispersion on the black hole mass and the AGN luminosity, in §3 we compare our predictions with observations of two samples of nearby weakly active galaxies. Some additional implications are described in §4, and the main conclusions are summarized in §5. An $H_0 = 80$ km s$^{-1}$ Mpc$^{-1}$ is assumed below, consistent with Tremaine et al. (2002).

## 2. THE LINE WIDTH LUMINOSITY RELATION

### 2.1. *The BLR*

There is now ample observational evidence that the BLR line widths, in particular the Balmer lines, are driven by gravity (e.g. Peterson, & Wandel 2000). The black hole mass, $M_{\rm BH}$, in AGN can thus be estimated using the H$\beta$ FWHM, $\Delta v$, and the size of the H$\beta$ emitting region, $R_{\rm BLR}$, obtained from the BLR size-luminosity relation (e.g., Kaspi et al. 2000). Assuming Keplerian motion gives $M_{\rm BH}({\rm H}\beta) = \Delta v^2 R_{\rm BLR}/G$, or

$$M_{\rm BH} = 10^{-21.7} \Delta v^2 L_{\rm bol}^{1/2} \ M_\odot, \qquad (1)$$

where $\Delta v$ is in km s$^{-1}$, and $L_{\rm bol}$ is the bolometric luminosity in erg s$^{-1}$ (e.g. Laor 1998). Using the definition of the Eddington luminosity, $L_{\rm Edd} \equiv 10^{38.1} M_{\rm BH}$ erg s$^{-1}$, the above relation implies

$$\dot{m} = 10^{-16.4} \Delta v^{-2} L_{\rm bol}^{1/2}, \qquad (2)$$

where $\dot{m} \equiv L_{\rm bol}/L_{\rm Edd}$[1]. Recent studies have shown that $M_{\rm BH}$ is correlated with the host bulge luminosity and with the host bulge stellar velocity dispersion, $\sigma_*$, as found for nearby non-active galaxies (Laor 1998; Gebhardt et al. 2000; Ferrarese et al. 2001), indicating that $M_{\rm BH}({\rm H}\beta)$ provides an estimate of the true $M_{\rm BH}$ to within a factor of 2–3.

The BLR line width is therefore related to the fundamental parameters, $M_{\rm BH}$, $\dot{m}$, and $L_{\rm bol}$ through

$$\Delta v = 21.1 M_{\rm BH}^{1/4} \dot{m}^{-1/4} \qquad (3)$$

and

$$\Delta v = 10^{10.85} M_{\rm BH}^{1/2} L_{\rm bol}^{-1/4}, \qquad (4)$$

implying that $\Delta v$ increases as $M_{\rm BH}$ increases and $\dot{m}$ and $L_{\rm bol}$ decrease. *How large can $\Delta v$ get?* Observations indicate the following; Forster et al. (2001) studied the H$\alpha$ emission line profiles of 141 AGN from the Large Bright Quasar Survey and found a maximum FWHM of 15,000 km s$^{-1}$, a similar maximum is seen by McIntosh et al. (1999) for H$\beta$ in a sample of 32 luminous AGN at $2.0 \leq z \leq 2.5$. Corbin (1997) studied the H$\beta$ profile of 45 radio-loud quasars, and found a maximum of 17,000 km s$^{-1}$, while Brotherton (1996) found a maximum 20,000 km s$^{-1}$ in a similar sample of 41 radio-loud quasars. Finally, Eracleous & Halpern (1994) found a maximum of 23,200 km s$^{-1}$ for the H$\alpha$ profile in a sample of 94 radio-loud AGN.

Eracleous & Halpern (1994) find that a large fraction of the very broad Balmer lines show a double peaked profile, unlike "normal", lower width, Balmer lines. This may suggest that broad double peaked lines originate in a different component (e.g. an accretion disk), which may not follow the relations used for deriving equation 1. However, there are some hints that double peaked Balmer lines originate from a component similar to lower width Balmer lines. First, reverberation mapping of 3C 390.3 indicates that its broad double peaked Balmer lines follow the same $R_{\rm BLR}$ vs. $L_{\rm bol}$ relation of lower luminosity objects (Dietrich et al. 1998; Kaspi et al. 2000). Second, normal width Balmer lines show indications of a strongly blended "hidden" double peaked component in their line profile variability characteristics (Wanders & Peterson 1996). Third, a "hidden" double peaked component is also indicated based on the polarization properties of some normal width objects (Smith et al. 2002). Thus, a double component structure may be a common property of Balmer lines, but since the two components are broad they are strongly blended in normal AGN, and become distinct only when the total velocity dispersion in the BLR is very large, as Eracleous & Halpern (1994) find. Below we assume that the BLR in objects with double peaked Balmer lines follows the same relations as in objects with normal width lines.

*Why are no objects with $\Delta v > 23,000$ km s$^{-1}$ observed?* Are broader lines impossible to detect? At $\Delta v = 30,000$ km s$^{-1}$ the H$\beta$ line width will be $\sim 500$ Å, and since the typical equivalent width of the broad H$\beta$ is $\sim 100$ Å, the mean line flux will be about 20% of the local continuum flux, which should be clearly detectable in reasonable quality spectra (S/N $\gtrsim 10$). Thus, the lack of detection of Balmer lines with $\Delta v > 23,000$ km s$^{-1}$, is probably not due to a low contrast effect, but rather appears to reflect the rarity/lack of very highly broadened Balmer line AGN. This lack may result from a physical upper limit on the velocity dispersion at which the BLR 'clouds' can survive, e.g. due to interactions with a surrounding medium, or due to large velocity shears and large tidal forces within the BLR.

It is important to note that in very low luminosity AGN the AGN continuum may be strongly diluted by the host galaxy light. As a result, the H$\beta$ equivalent width will be significantly reduced, and a very broad H$\beta$ may be undetectable even in high quality spectra. Thus, there is currently no good observational constraint on the maximum $\Delta v$ in very low luminosity AGN. The detection of broad lines in such objects requires high spatial resolution spectroscopy which can reject most of the host galaxy light (some examples are mentioned in §3).

If the BLR "clouds" can survive only at $\Delta v \leq 25,000$ km s$^{-1}$, then eqs. 3 & 4 imply a minimum luminosity

$$L_{\rm min} = 10^{25.8} M_{\rm BH}^2, \qquad (5)$$

or equivalently a minimum accretion rate,

$$\dot{m}_{\rm min} = 10^{-12.3} M_{\rm BH}, \qquad (6)$$

below which $\Delta v > 25,000$ km s$^{-1}$, and the BLR will not survive. However, the narrow lines will still be present, and such objects will constitute "true" type 2 AGN, i.e. they will not show any evidence for an obscured BLR. Apart from having $L_{\rm bol} < L_{\rm min}$, these "true" type 2 AGN may be otherwise normal, i.e. the lack of a BLR would not necessarily imply a significantly different inner structure compared to normal AGN.

### 2.2. *The NLR*

What happens to the velocity dispersion at the Narrow Line Region (NLR) when $\dot{m}$ or $L_{\rm bol}$ are very low? In normal luminous AGN the NLR is far enough from the center to have its dynamics dominated by the bulge, as indicated by the correlation of $\sigma_*$ and the [O III]$\lambda\lambda4959, 5007$ line width (Nelson & Whittle 1996). However, when $\dot{m}$ or $L_{\rm bol}$ are very low the inner part of the NLR may contract enough to enter the realm of the black hole. To estimate the minimum possible size of the [O III]$\lambda\lambda4959, 5007$ emitting region, $R_{\rm [OIII]}$, we assume an electron density at the [O III] emitting region in the NLR of the order of the critical density, $n_{\rm NLR} \sim n_{\rm crit} \sim 10^6$ cm$^{-3}$, which is $\sim 10^{-4} n_{\rm BLR}$. The NLR ionization parameter, defined as $U_{\rm NLR} \equiv n_\gamma/n_{\rm NLR}$, where $n_\gamma$ is the H ionizing photon density at the NLR, is about $\sim 10^{-3}$ (below that [O III] becomes an inefficient coolant), which is $\sim 10^{-2} U_{\rm BLR}$. Since generally $R \propto n_\gamma^{-1/2} \propto (nU)^{-1/2}$, we get that $R_{\rm [OIII]} \sim 10^3 R_{\rm BLR}$. Thus, since $\Delta v \propto R^{-1/2}$ in a Keplerian velocity field, the black hole will produce a velocity dispersion at the NLR of $\sigma_{\rm [OIII]}^{\rm BH} = \sigma_{\rm BLR}/31.6$ or

$$\sigma_{\rm [OIII]}^{\rm BH} = 0.28 M_{\rm BH}^{1/4} \dot{m}^{-1/4}, \qquad (7)$$

using eq. 3 and assuming $\sigma = \Delta v/2.35$ (appropriate for a Gaussian line profile). The bulge velocity dispersion, $\sigma_*$, is related to $M_{\rm BH}$ through the Tremaine et al. (2002) relation

$$\sigma_* \simeq 1.9 M_{\rm BH}^{1/4}, \qquad (8)$$

---

[1] This relation is one of the motivations for the suggestion that narrow line AGN are shining at $\dot{m} \sim 1$ (e.g. Laor 2000 and references therein).

and thus we get the ratio of the black hole to the bulge contributions to the velocity dispersion at the innermost NLR

$$\sigma_{\rm [OIII]}^{\rm BH}/\sigma_* = 0.15 \dot{m}^{-1/4}. \qquad (9)$$

This ratio is only a function of $\dot{m}$, and thus this inner NLR is dominated by the black hole, i.e. $\sigma_{\rm [OIII]}^{\rm BH}/\sigma_* > 1$, once $\dot{m} < 5 \times 10^{-4}$. Narrow lines of lower $n_{\rm crit}$ will be produced at larger distances and thus require lower $\dot{m}$ to be dominated by the BH. E.g. for [N II]$\lambda\lambda 6548, 6583$ $n_{\rm crit} \simeq 10^5$, and the transition from bulge dominated to black hole dominated dynamics occurs at $\dot{m} < 5 \times 10^{-5}$. Such a transition was apparently observed by Filippenko & Sargent (1988), as further discussed in §4.1. Significant narrow line flux may also be produced on much larger scales, as recently found by Bennert et al. (2002) for the [O III] line.

### 3. COMPARISON WITH OBSERVATIONS

To test the validity of the proposed explanation for the lack of a BLR in low luminosity objects we compare below the expected and observed Balmer line widths in two samples of nearby AGN dominated by low luminosity objects. This allow us to predict which of the type 2 objects in these samples may turn out to have an obscured BLR, and which are likely to be true type 2 AGN.

The first sample is based on the Palomar survey for "dwarf Seyfert nuclei" (Ho, Filippenko, & Sargent 1995). This sample includes $\sim 500$ objects, and we therefore restrict our sample to the subsample of objects with apparently broad H$\alpha$ listed in Table 1 of Ho, et al. (1997a). For each object we compare the predicted and observed $\Delta v$ and conclude whether it is likely to have a hidden BLR, or not.

#### 3.1. How to estimate $\Delta v$

To estimate $\Delta v$ we need to estimate $L_{\rm bol}$ and $M_{\rm BH}$. The bolometric luminosity is estimated from the $B$ band nuclear magnitude, $M_B^{\rm nuc}$ measured by Ho & Peng (2001), using the standard calibration

$$\log \nu L_\nu(B) = 35.568 - 0.4 M_B^{\rm nuc}, \qquad (10)$$

together with the Ho (1999) estimate of the mean bolometric correction factor

$$L_{\rm bol} = 24 \nu L_\nu(B) \qquad (11)$$

for low luminosity AGN. An additional way to estimate $L_{\rm bol}$ is through the narrow H$\alpha$ line luminosity, $L({\rm H}\alpha)$, available for a very large number of nearby galaxies in Ho, Filippenko, & Sargent (1997b) and Ho, Filippenko, & Sargent (2003)[2]. We use the Ho & Peng relation

$$\log L({\rm H}\beta) = -0.34 M_B^{\rm nuc} + 35.1 \qquad (12)$$

and the assumption $L({\rm H}\alpha)/L({\rm H}\beta)=3$ to estimate $M_B^{\rm nuc}$ from $L({\rm H}\alpha)$, which gives

$$\log L_{\rm bol} = 1.176 \log L({\rm H}\alpha) - 4.91. \qquad (13)$$

In some objects both $M_B^{\rm nuc}$ and $L({\rm H}\alpha)$ are available, and they may imply somewhat different values for $L_{\rm bol}$. In such cases we use the higher estimate of the two for $L_{\rm bol}$. This helps minimize effects of extinction on either $M_B^{\rm nuc}$ or $L({\rm H}\alpha)$, though it may overestimate $L_{\rm bol}$ if H$\alpha$ is mostly produced by stellar photoionization rather than by a weak AGN.

The black hole mass is estimated using $\sigma_*$ and the Tremaine et al. (2002) relation,

$$\log M_{\rm BH} = -1.12 + 4.02 \log \sigma_*, \qquad (14)$$

where $\sigma_*$ is obtained from the comprehensive studies of McElroy (1995) and Prugniel & Simien (1996), supplemented by the AGN study of Nelson & Whittle (1995). We also use some recent results from Ferrarese et al. (2001), Sarzi et al. (2001), and Barth, Ho & Sargent (2002). These studies yield $\sigma_*$ values for 35 of the 46 AGN in the Ho et al. sample, which are listed in Table 1 here (excluding three objects which are listed in Table 2, see below). These studies generally give consistent $\sigma_*$ for a given object, but in four objects the different best estimates for $\sigma_*$ differ by $> 20\%$, and in these cases the two extreme values are listed and used in Table 1.

Thus, given $\sigma_*$ and $L_{\rm bol}$, and using eqs. 4 & 14, the predicted $\Delta v$ is given by

$$\log \Delta v_{\rm pred} = 10.29 + 2.01 \log \sigma_* - 0.25 \log L_{\rm bol}, \qquad (15)$$

and the predicted $\dot{m}$ is

$$\log \dot{m} = -36.977 + \log L_{\rm bol} - 4.02 \log \sigma_*. \qquad (16)$$

These two parameters are listed in Table 1 for each object, together with the observed[3] $\Delta v$.

#### 3.2. The accuracy of the $\Delta v$ estimate

The above estimate of $\Delta v$ is based on a number of observed relations, each of which has significant scatter, and the cumulative scatter may be too large to make this estimate useful. Below we test the accuracy of the $\Delta v$ estimate by comparing it with the observed $\Delta v$ in objects where broad Balmer lines are clearly observed.

Three AGN in the Ho et al. sample are classified as S1.0 or S1.2 (NGC 3516, NGC 4051, NGC 4639). This classification implies that the BLR of these objects is directly observed. The observed and predicted $\Delta v$ of these three objects (Table 1) agree to within a factor of 2–3. There are also seven S1.5 objects in this sample. These AGN have weaker broad lines which may imply that part of the BLR is not detected. In four of the S1.5 objects the observed and predicted $\Delta v$ also agree to a factor of 2–3, and in two other (NGC 1275, NGC 5033) the discrepancy is by a factor of $\sim 3.5$ (but see footnote $c$ in Table 1). In the remaining S1.5 galaxy, NGC 3031, there is a large discrepancy ($\Delta v_{\rm pred} = 22,000$ km s$^{-1}$ vs. $\Delta v_{\rm obs} = 2650$ km s$^{-1}$), but *Hubble Space Telescope* (*HST*) spectroscopy by Bower et al. (1996) reveals a very broad double peaked base component ($\Delta v_{\rm obs} \simeq 10,000$), which may be the true BLR of this object. Thus, we conclude that the method outlined above for estimating the Balmer line widths based on $\sigma_*$ and either $L({\rm H}\alpha)$ or $M_B^{\rm nuc}$, is accurate to about a factor of 2–3 in most cases.

Interestingly, the factor of 2–3 discrepancy is systematic, with $\Delta v_{\rm pred} > \Delta v_{\rm obs}$ in all cases. This may partly result from extinction, as indicated by $L({\rm H}\alpha)/L({\rm H}\beta) > 3$ in some of the objects (Ho et al. 1997b), which would lead to an underestimate of $L_{bol}$, and thus overestimate of $\Delta v_{pred}$. In addition, the value of $\Delta v_{obs}$ may be underestimated due to the low S/N at the base of the Balmer line profiles in ground based spectra, as most objects in Table 1 are strongly dominated by the host galaxy flux (further evidence for that is discussed below). More accurate values of $L_{bol}$ and $\Delta v_{obs}$ are required to asses whether a better agreement between $\Delta v_{pred}$ and $\Delta v_{obs}$ can be achieved.

---

[2] Some of the distances in Ho et al. 1997b are based on $H_0 = 75$ km s$^{-1}$ Mpc$^{-1}$, which results in a negligible error compared to the overall uncertainties described below

[3] Note that $\Delta v_{\rm pred}$ is obtained for H$\beta$, while the observations in Table 1 are for H$\alpha$, which is typically observed to be $\sim 10\%$ narrower than H$\beta$.

### 3.3. *The observed vs. predicted $\Delta v$ in type 1.8-2 objects*

There are seven S1.9/L1.9 class objects in Table 1 where $\Delta v_{pred} = 12,000 - 22,000$ km s$^{-1}$, and $\Delta v_{obs} = 1500 - 3000$ km s$^{-1}$. These may be good candidates for true type 1 AGN. Indeed, high spatial resolution spectroscopy with *HST* reveals $\Delta v_{obs} \sim 8000-9500$ km s$^{-1}$ components to the H$\alpha$ lines in three of these objects (NGC 4203, Shields et al. 2000; NGC 4450, Ho et al. 2000; NGC 4579, Barth et al. 2001a), which are within a factor of $\sim 2$ of $\Delta v_{pred}$ for those objects. The remaining four objects (NGC 2639, NGC 2681, NGC 3642, NGC 4565) may reveal similar broad components in high quality *HST* spectroscopy, if their BLR is not obscured. Ground based spectropolarimetry of two of these objects by Barth et al. (1999; NGC 2639, NGC 3642) does not reveal a hidden polarized broad line component, but the S/N may not be large enough in these ground based spectra.

There are also 12 S1.9/L1.9/T1.9 class objects in Table 1 with $\Delta v_{pred} = 31,000 - 120,000$ km s$^{-1}$. We suggest that most of these are likely to prove to be true type 2 AGN. The spectropolarimetric survey of Barth et al. (1999) includes three of these 12 objects. In NGC 315 it reveals a polarized slightly broadened [N II]+H$\alpha$ blend, in NGC 1052 a $\Delta v_{obs} = 5000$ km s$^{-1}$ H$\alpha$ line is seen in scattered light, and in NGC 3998 no polarization is detected. High spatial resolution spectroscopy with *HST* can reveal if these objects are not true type 2 AGN.

An additional qualitatively different sample is presented in Table 2, which includes mostly very weak AGN. This sample is based on the Tremaine et al. (2002) study which provides a list of the best $M_{BH}$ estimates for 31 galaxies (Table 1 there). These are generally nearby normal galaxies with no signs of nuclear activity. However, 10 of these objects have a generally very weak nuclear $L$(H$\alpha$), as measured by Ho et al. (1997b), which allows us to estimate their $L_{bol}$ (eq. 13), as listed in Table 2. Since a direct estimate of $M_{BH}$ is available for all these objects, there is no need to use the $\sigma_*$ based relations to obtain their $\Delta v_{pred}$ and $\dot{m}$. Instead, we use the $M_{BH}$, $L_{bol}$ based relations

$$\log \Delta v = 10.85 + 0.5 \log M_{BH} - 0.25 \log L_{bol} \qquad (17)$$

and

$$\log \dot{m} = -38.1 + \log L_{bol} - \log M_{BH}. \qquad (18)$$

These two parameters are also listed in Table 2. There are three objects which belong to the Ho et al. sample as well (NGC 1068, NGC 2787, NGC 4258), and these are listed in Table 2. Excluding NGC 1068, all other nine objects in this sample have $\Delta v_{pred} \geq 38,000$ km s$^{-1}$, and are thus likely to be true type 2 AGN.

Table 2 includes NGC 4486, where diffraction limited ($\sim 0.''3$) mid IR spectroscopy by Whysong & Antonucci (2002) indicates no significant thermal emission component from the nucleus. Such a component should have been present near the nucleus if an obscuring material with a non-negligible covering factor was present. As argued by Whysong & Antonucci, the absence of such a thermal component provides strong evidence that NGC 4486 is a true type 2 AGN, as also indicated by its $\Delta v_{pred} = 169,000$ km s$^{-1}$. The high spatial resolution *HST* spectroscopy of NGC 4486 by Sabra et al. (2003) shows no broad optical or UV lines, and thus indicates that the lack of broad lines is not due to host galaxy light contamination.

Figure 1 displays the distribution of the two samples of objects in the $\sigma_*$ vs. $L_{bol}$ plane. The diagonal lines mark the position of objects with a given $\Delta v_{pred}$, or a given $\dot{m}$. This figure allows a quick assessment if an object with a given $\sigma_*$ and $L_{bol}$ is likely to be a so called "narrow line seyfert 1" (NLS1[4]), whether it is likely to show very broad Balmer lines, which are often double peaked, as seen in many nearby radio galaxies (Eracleous & Halpern 1994), or whether it is likely to be a true type 2 AGN.

## 4. ADDITIONAL IMPLICATIONS

### 4.1. *The Narrow Line Widths*

Although all the objects in Table 1 were classified by Ho et al. (1997a) as having a broad , i.e. $\Delta v_{obs} \sim 2000 - 3000$ km s$^{-1}$ component in H$\alpha$, this weak broad component may in some cases be an artifact of the narrow line profile decomposition scheme. Small residual flux which remains from the profile fits to the [N II]$\lambda\lambda 6548,6583$ doublet components (separated by $\sim 1600$ km s$^{-1}$) may be interpreted as a single broad component centered at H$\alpha$. This problem can be particularly significant in objects with large $\sigma_*$, as these tend to have a relatively broad and asymmetric base to their forbidden lines. An additional interesting possibility is a broad forbidden line component arising from a compact NLR. As mentioned in §2.2, the innermost NLR may extend down to $\sim 10^3 R_{BLR}$, which implies some narrow line emission from gas with $\Delta v_{NLR} \sim \Delta v/30$. Thus, relatively broad bases may be present in the narrow line profiles of objects where $\Delta v_{pred} \gtrsim 60,000$ km s$^{-1}$, such as NGC 315, NGC 1161, NGC 3998, NGC 4168 and NGC 4636 (see Table 1), and a BLR H$\alpha$ component may not be required in these objects.

Nine of the ten objects in the Tremaine et al. sample (excluding NGC 1068, see Table 2), have very low $\dot{m}$, generally well below $10^{-5}$. This indicates that some of the NLR may be dynamically dominated by the BH (see §2). High spatial resolution spectroscopy with HST shows this effect very clearly in NGC 4486 (Harms et al. 1994; Macchetto et al. 1997), and to a somewhat lesser extent in NGC 4261 (Ferrarese, Ford, & Jaffe 1996), NGC 3245 (Barth et al. 2001b), and NGC 2787, NGC 4459, and NGC 4596 (Sarzi et al. 2001).

Various studies find strong dependence of the narrow line width on $n_{crit}$ and/or ionization potential, IP, (e.g. Filippenko & Halpern 1984; Filippenko 1985), indicating density stratification with distance in the NLR, as directly confirmed by *HST* imaging in some cases (Barth et al. 2001a). However, other detailed studies do not always find such a clear dependence (e.g. De Robertis & Osterbrock 1984, 1986; Whittle, 1985; Veilleux 1991). These apparently conflicting results may be related to the dependence of $\sigma^{BH}/\sigma_*$ on $\dot{m}$, as discussed in §2.2. For example, there is a clear dependence of $n_{crit}$ on line width in NGC 7213 (Filippenko & Halpern 1984), where $\sigma_* = 185$ km s$^{-1}$ (Nelson & Whittle 1985), and $L$(H$\alpha$)$= 10^{40.56}$ (narrow component in Table 2 of Filippenko & Halpern, and a distance of 19.8 Mpc), implying that $\dot{m} = 5 \times 10^{-4}$. Thus, lines with $n_{\text{crit}} \gtrsim 10^6$ cm$^{-3}$ in NGC 7213 can be dominated by the black hole potential. A particularly striking example is available in NGC 3031 (M 81), where Filippenko & Sargent (1988, fig. 6 there) find a strong dependence of the narrow line width on $n_{\text{crit}}$ for lines with $n_{\text{crit}} > 10^5$ cm$^{-3}$, and a constant line width for lines with $n_{\text{crit}} < 10^5$ cm$^{-3}$. They interpret this as the transition from the nearly constant bulge velocity field to the rising velocity dispersion with decreasing distance within the realm of the BH. According to our estimates, $\dot{m} = 5.4 \times 10^{-5}$ in NGC 3031 (Table 1), and at this value of $\dot{m}$ the transition from BH to bulge dominance should occur at $n_{\text{crit}} = 10^5$ cm$^{-3}$ (see §2.2), as observed. This provides a surprisingly good quantitative support to the interpretation of Filippenko & Sargent (1988).

---

[4] NLS1 are generally defined as having $\Delta v < 2000$ km s$^{-1}$. Note that being a NLS1 and having $\dot{m} \sim 1$ are not exactly overlapping properties, as further discussed in Laor (2000, §7 there).

In objects with $\dot{m} \gtrsim 10^{-3}$ all the NLR should reside in the bulge potential. This is consistent with the correlation of $\sigma_*$ and the [O III] line width found by Nelson & Whittle (1996) in Seyfert galaxies, and further supported by the recent studies of Nelson (2000) and Shields et al. (2003). However, at a high enough $\dot{m}$ radiation pressure may drive strong outflow in the NLR (e.g. as observed in NGC 1068 by Cecil et al. 2002), which can also lead to a radial velocity stratification with distance. The radiation pressure induced velocity field may not be single valued, and it can be more complicated than the simple Keplerian dependence expected at very low $\dot{m}$. Establishing the importance of the $\dot{m}$ effect for the narrow line width vs. $n_{\text{crit}}$ and IP relations requires a detailed study of the NLR profiles in AGN spanning a wide range of $\dot{m}$ values.

### 4.2. *Other low luminosity type 2 AGN*

Low luminosity AGN tend to be of type 2. This tendency is clearest for radio loud AGN. Specifically, FR I radio galaxies[5] are generally of low luminosity, and are almost exclusively type 2 AGN (Baum, Zirbel & O'Dea 1995), while FR II radio galaxies are generally more luminous, and are sometimes of type 1. In particular, Chiaberge, Capetti & Celotti (2002a) find that all FR II galaxies above an optical luminosity of $L_{\text{opt}} = 4 \times 10^{42}$ erg s$^{-1}$ are of type 1. Baum et al. (1995), Chiaberge, Capetti, & Celotti (1999), and Chiaberge et al. (2002b) discuss various evidence indicating that the core of FR I galaxies is unobscured, and thus a hidden BLR is unlikely to be present. A similar conclusion is reached below based on the $\Delta v(M_{\text{BH}}, L_{\text{bol}})$ relation.

The radio power of FR I and FR II galaxies is generally above $10^{40}$ erg s$^{-1}$ (e.g. Baum et al. 1995), which suggests they typically have $M_{\text{BH}} \gtrsim 10^{7.5} M_\odot$ (Laor 2000, fig.3; Lacy et al. 2001, fig.2; Ho 2002, fig.2; Jarvis & McLure 2002, fig.3; c.f. Oshlack et al. 2002, fig.5; Woo & Urry 2002, fig.10). This limit implies $\log \sigma_* \gtrsim 2.15$. The low line luminosity, $L(\text{H}\alpha) \lesssim 10^{41}$ erg s$^{-1}$, of FR I galaxies (Baum et al. 1995) suggests $L_{\text{bol}} \lesssim 10^{43}$ erg s$^{-1}$, as also indicated by their core optical luminosities measured with *HST* (Chiaberge et al. 2002b). This range of parameters, $\log \sigma_* \gtrsim 2.15$ and $L_{\text{bol}} \lesssim 10^{43}$ erg s$^{-1}$, suggests that most FR I galaxies would be true type 2 AGN (see Fig.1). Similarly, the finding of Chiaberge et al. (2002a) that all FR II galaxies with $L_{\text{bol}} \gtrsim 4 \times 10^{43}$ erg s$^{-1}$ ($L_{\text{bol}} \sim 10 L_{\text{opt}}$) are of type 1, is consistent with their having $\Delta v_{\text{pred}} < 25,000$ km s$^{-1}$ (see Fig.1).

As stressed by Antonucci (2002), there are a few known cases of broad line FR I galaxies, such as 3C 120, B2 1028+313, E1821+643, and J2114+820. Such objects appear to contradict the suggestion that FR I galaxies are the parent population of BL Lac objects, where no broad lines are present (e.g. Lara et al. 1999). However, these four objects are also unusually luminous for FR I galaxies, all having $L_{\text{bol}} \gtrsim 5 \times 10^{44}$ erg s$^{-1}$. At this luminosity range a BLR can be present, and this provides additional support to the suggestion that typical FR I galaxies lack a BLR because of their low luminosity.

Recent spectropolarimetric surveys which look for hidden broad line regions in type 2 AGN show a lower detection rate of hidden BLR in lower luminosity AGN. This appears to be a real effect, i.e. it is not driven by a S/N effect (Tran 2001, 2003; but see Antonucci 2001; Gu & Huang 2002), suggesting that true type 2 AGN are more common at low luminosity. Similarly, Panessa & Bassani (2002) find an increasing fraction of type 2 AGN without significant X-ray absorption, with decreasing luminosity[6]. Both results are qualitatively consistent with the $\Delta v(M_{\text{BH}}, L_{\text{bol}})$ relation discussed above. It will be interesting to explore quantitatively if the suggested true type 2 AGN and the hidden type 1 AGN have significantly different predicted $\Delta v$.

It is important to note that the lack of detection of broad lines in some low luminosity type 2 AGN may also be due to a combination of host galaxy dilution together with a very large $\Delta v$. High spatial resolution *HST* spectra can establish if a very broad BLR component does exist in low luminosity type 2 AGN.

### 4.3. *What sets the size of the BLR?*

The $M_{\text{BH}}$ estimate of §2.1 uses the relation $R_{\text{BLR}} = 0.086(L_{\text{bol}}/10^{46})^{1/2}$ pc (Laor 1998). Observations suggest a steeper slope, $\sim 0.6 - 0.7$, when the monochromatic luminosity at 5100 Å, $L_{5100}$, is used (kaspi et al. 2000). However, the 0.5 slope used here may still be viable since the bolometric correction factor $L_{\text{bol}}/L_{5100}$ increases with luminosity. The 0.5 slope is appealing since dust can strongly suppress line emission (Laor & Draine 1993; Netzer & Laor 1993; Ferguson et al. 1997), and thus its complete sublimation, which occurs at $R_{\text{sub}} \simeq 0.2(L_{\text{bol}}/10^{46})^{1/2}$ pc, provides a natural, parameter free, explanation for the outer boundary of the BLR, which is consistent with the $R_{\text{BLR}}$ relation used above. The inner boundary of the BLR may be set by thermal quenching of the line emission (Ferland & Rees 1988; Rees, Netzer, & Ferland 1989). In this scenario the BLR does not mark the location of photoionized gas near AGN, but rather the location where this gas cools mainly by line emission, rather than by either thermal dust emission, just outside the BLR, or by free-free emission inside the BLR (Phinney 1989).

Nicastro (2000) suggested that the BLR is formed by an outflow from an accretion disk. An outflow which occurs in the region bounded on the inward side by $R_{\text{tran}}$, the transition radius between the radiation pressure and gas pressure dominated region of the disk, and on the outside by the largest radius which presumably allows an outflowing confining corona. In this scenario $R_{\text{BLR}}/M_{\text{BH}} \propto \dot{m}^{0.66-0.76}$, i.e. $R_{\text{BLR}}$ is a function of both $L_{\text{bol}}$ and $M_{\text{BH}}$. This scenario may be inconsistent with the observations which suggest dependence on $L_{\text{bol}}$ only (e.g. Kaspi et al. 2000). Interestingly, this model predicts that at $\dot{m} \lesssim \dot{m}_{\text{min}} = (1-4) \times 10^{-3}$ no BLR will be present since the disk is gas pressure dominated throughout (Shakura & Sunyaev 1973). The analysis presented here suggests a BLR can be present down to $\dot{m} \sim 10^{-4}$ (e.g. NGC 3031 and NGC 4639, Table 1), but given our poor understanding of accretion disks, this discrepancy may not rule out the proposed mechanism. If there indeed exists a critical $\dot{m}$, that would imply $\Delta v_{max} \propto L_{bol}^{1/4}$ (e.g. eq. 2), i.e. the maximum possible Balmer line width in AGN increases with luminosity, and is not a constant as assumed here. It is therefore interesting to establish observationally whether the critical parameter which controls the existence of a BLR is $\Delta v$, as assumed here, or $\dot{m}$, as proposed by Nicastro.

---

[5] FR I radio galaxies are defined as having "edge darkened" radio morphology, and "edge brightened" radio morphology defines FR II radio galaxies (Fanaroff & Riley 1974).

[6] Note that some of the Seyfert 2 AGN in Panessa & Bassani, e.g. NGC 4579, are actually classified as type 1.9, and the weakness of their BLR may be due to stellar light dilution, rather than obscuration of the central parts (Barth 2002, §2.2 there).

Another origin for the BLR was proposed by Collin & Hure (2001), who suggested the BLR clouds forms where the disk becomes gravitationally unstable. They find that this scenario implies $R_{BLR} \propto M_{BH}^{0.54}$, using the Shakura & Sunyaev disk solution, i.e. $R_{BLR}$ is independent of luminosity. The observed strong dependence of $R_{BLR}$ on $L_{bol}$ requires an additional mechanism which determines whether the illuminated gas clouds produce broad lines. Since all accretion disks must become gravitationally unstable far enough from the center, this mechanism does not provide a natural explanation for the apparent absence of a BLR in some AGN.

We note in passing the systematic change in the SED of some AGN at very low values of $L_{bol}$ (e.g. Eracleous & Halpern 1994; Baum et al. 1995; Ho 1999), which may be driven by a change in the accretion configuration at very low $\dot{m}$, as predicted by some AGN accretion models (e.g. Rees et al. 1982). If this change in accretion configuration extends to a large enough radius, it may also affect the formation of the BLR, e.g. if the BLR forms in a disk wind (Shlosman, Vitello, & Shaviv, 1985; Emmering, Blandford, & Shlosman, 1992; Murray et al. 1995). This line of speculations may also lead to $\dot{m}$ as the primary parameter controlling the existence of a BLR (e.g. Chiaberge et al. 2002a).

Another possible speculation is that the BLR is disrupted above a certain tidal field (Scoville 2002), which is $\propto (\Delta v / R_{BLR})^2$, or above a certain radial velocity shear, $\propto \Delta v / R_{BLR}$. Both mechanisms imply $L_{min} \propto M_{BH}^{2/3}$, a critical accretion rate $\dot{m}_{min} \propto M_{BH}^{-1/3}$, and $\Delta v_{max} \propto L_{bol}^{1/2}$. Such a strong dependence of $\Delta v_{max}$ on $L_{bol}$ may not be consistent with the observed large $\Delta v$ in both high and low luminosity AGN. However, more systematic studies of $\Delta v_{max}$ vs. $L_{bol}$ are required for a more definite conclusion.

It is important to note that the absence of a BLR in some objects cannot be just due to very low values of $L_{bol}$, as there are clear cases of very low luminosity AGN with a BLR (e.g. the *HST* detections in NGC 4203, NGC 4450, see Table 1). The most striking example is NGC 4395 (Filippenko, Ho, & Sargent 1993), a type 1 AGN at $L_{bol} \simeq 1.9 \times 10^{40}$ erg s$^{-1}$ (Moran et al. 1999). In this object $M_{BH} < 8 \times 10^{4} M_{\odot}$ (Moran et al. 1999), and eq. 4 implies $\Delta v_{pred} < 1700$ km s$^{-1}$, consistent with $\Delta v_{obs} \sim 1500$ km s$^{-1}$ (Kraemer et al. 1999). Thus, the crucial parameter for the presence of a BLR is most likely a low $\dot{m}$, rather than just a low $L_{bol}$. The critical $\dot{m}$ may be proportional to $M_{BH}$, as proposed in §2 (eq. 6), to $M_{BH}^{-1/3}$ as mentioned above, or it may be a universal constant, as some disk models may suggest.

## 5. CONCLUSIONS

The purpose of this paper is to point out a possible physical reason for the apparent absence of broad lines in most low luminosity AGN. The size of the BLR may be limited from the outside by the dust sublimation radius, which can set the observed $R_{BLR}$ vs. $L_{bol}$ relation. In addition, if the BLR cannot survive once $\Delta v \gtrsim 25,000$ km s$^{-1}$, then when $L_{bol} < L_{min} = 10^{41.8}(M_{BH}/10^{8}M_{\odot})^2$, or equivalently when $\dot{m} < 10^{-4.3}M_{BH}/10^{8}M_{\odot}$, $R_{BLR}$ shrinks below the radius where $\Delta v = 25,000$ km s$^{-1}$, leaving no region where the BLR can survive. As a result, AGN with $L_{bol} < L_{min}$ will not have a BLR, and they will form a class of true type 2 AGN. These true type 2 AGN do not contradict the AGN inclination unification schemes, and they may be a natural outcome of the processes which determine the inner and outer boundary of the BLR.

A simplified method to estimate $\Delta v$ based on $\sigma_*$ and $L_{bol}$ (derived from either $M_B^{nuc}$ or $L(H\alpha)$) is described (§3.1), its accuracy is measured (§3.2), and it is used for providing $\Delta v_{pred}$ for a set of nearby low luminosity AGN (§3.3). These predictions can be tested with future higher quality observations. We suggest that most objects with $\Delta v_{pred} > 25,000$ km s$^{-1}$ listed in Tables 1 & 2, are likely to be true type 2 AGN, i.e. they will not reveal a hidden BLR in either spectropolarimetry, or IR spectroscopy. The set of AGN with $\Delta v_{pred} =10,000–25,000$ km s$^{-1}$ is likely to reveal very broad Balmer lines in high spatial resolution spectroscopy, if their BLR is not obscured. It will be interesting to explore whether the absence of broad lines in low luminosity FR I galaxies, and other low luminosity type 2 AGN, is indeed related to having $L_{bol} < L_{min}$.

The value of $\dot{m}$ may control the ratio of the BH/bulge potential at the NLR, and a Keplerian radial velocity stratification of the inner NLR may be observed in all objects with $\dot{m} < 5 \times 10^{-4}$.

The crucial assumption behind the $L_{min}$ conjecture for the existence of a BLR is that the BLR ceases to exist above some critical $\Delta v$. *Is there a real $\Delta v$ cutoff? If yes, is this cutoff universal, or is it a function of luminosity?* Extremely little is currently known about these issues, in particular in low luminosity AGN. Some answers can be obtained through carefull spectroscopic searches for weak broad Balmer lines (e.g. Barth et al. 2001a; Ho et al. 2000; Shields et al. 2000) in large well defined sample of AGN straddling either $L_{bol} \sim 10^{41.8}(M_{BH}/10^{8}M_{\odot})^2$, or possibly $\dot{m} \sim 10^{-4}$. The answers may bear important clues to the physical nature of the BLR.

Some very useful comments by Ski Antonucci, Aaron Barth, Dani Maoz, Joseph Shields, and the referee are gratefully acknowledged.

REFERENCES

Antonucci, R. R. J. 1993, ARA&A, 31, 473
Antonucci, R. 2001, IAU Colloq. 184: AGN Surveys, Eds. R.F. Green, E.Ye. Khachikian, and D.B. Sanders, (SF, ASP), 1 (astro-ph/0110343)
Antonucci, R. 2002, Astrophysical Spectropolarimetry, eds. J. Trujillo-Bueno, F. Moreno-Insertis, and F. Sánchez, (Cambridge, CUP), 151 (astro-ph/0103048)
Antonucci, R. R. J. & Miller, J. S. 1985, ApJ, 297, 621
Barth, A. J. 2002, In *Issues in Unification of AGNs*, eds. R. Maiolino, A. Marconi, & N. Nagar, ASP (astro-ph/0201065)
Barth, A. J., Filippenko, A. V., & Moran, E. C. 1999, ApJ, 525, 673
Barth, A. J., Ho, L. C., Filippenko, A. V., Rix, H., & Sargent, W. L. W. 2001a, ApJ, 546, 205
Barth, A. J., Sarzi, M., Rix, H., Ho, L. C., Filippenko, A. V., & Sargent, W. L. W. 2001b, ApJ, 555, 685
Barth, A. J., Ho, L. C., & Sargent, W. L. W. 2002, ApJ, 124, 2607

Baum, S. A., Zirbel, E. L., & O'Dea, C. P. 1995, ApJ, 451, 88
Bennert, N., Falcke, H., Schulz, H., Wilson, A. S., & Wills, B. J. 2002, ApJ, 574, L105
Brotherton, M. S. 1996, ApJS, 102, 1
Cecil, G., Dopita, M. A., Groves, B., Wilson, A. S., Ferruit, P., Pécontal, E., & Binette, L. 2002, ApJ, 568, 627
Chiaberge, M., Capetti, A., & Celotti, A. 1999, A&A, 349, 77
– 2002a, A&A, 394, 791
Chiaberge, M., Macchetto, F. D., Sparks, W. B., Capetti, A., Allen, M. G., & Martel, A. 2002b, ApJ, 571, 247
Collin, S. & Huré, J.-M. 2001, A&A, 372, 50
Corbin, M. R. 1997, ApJS, 113, 245
De Robertis, M. M., & Osterbrock, D. E. 1984, ApJ, 286, 171
– 1986, ApJ, 301, 727
Dietrich, M. et al. 1998, ApJS, 115, 185
Emmering, R. T., Blandford, R. D., & Shlosman, I. 1992, ApJ, 385, 460
Eracleous, M. & Halpern, J. P. 1994, ApJS, 90, 1
Fanaroff, B. L. & Riley, J. M. 1974, MNRAS, 167, 31P
Ferland, G. J. & Rees, M. J. 1988, ApJ, 332, 141
Ferguson, J. W., Korista, K. T., Baldwin, J. A., & Ferland, G. J. 1997, ApJ, 487, 122
Ferrarese, L., Ford, H. C., & Jaffe, W. 1996, ApJ, 470, 444
Ferrarese, L., Pogge, R. W., Peterson, B. M., Merritt, D., Wandel, A., & Joseph, C. L. 2001, ApJ, 555, L79
Filippenko, A. V. 1985, ApJ, 289, 475
Filippenko, A. V., & Halpern, J. P. 1984, ApJ, 285, 458
Filippenko, A. V., Ho, L. C., & Sargent, W. L. W. 1993, ApJ, 410, L75
Filippenko, A. V., & Sargent, W. L. W. 1988, ApJ, 324, 134
Forster, K., Green, P. J., Aldcroft, T. L., Vestergaard, M., Foltz, C. B., & Hewett, P. C. 2001, ApJS, 134, 35
Gallagher, S. C., Brandt, W. N., Laor, A., Elvis, M., Mathur, S., Wills, B. J., & Iyomoto, N. 2001, ApJ, 546, 795
Gebhardt, K. et al. 2000, ApJ, 543, L5
Green, P. J., Aldcroft, T. L., Mathur, S., Wilkes, B. J., & Elvis, M. 2001, ApJ, 558, 109
Gu, Q. & Huang, J. 2002, ApJ, 579, 205
Harms, R. J. et al. 1994, ApJ, 435, L35
Ho, L. C. 1999, ApJ, 516, 672
– 2002, ApJ, 564, 120
Ho, L. C., Filippenko, A. V., & Sargent, W. L. 1995, ApJS, 98, 477
– 1997b, ApJS, 112, 315
– 2003, ApJ, 583, 159
Ho, L. C., Filippenko, A. V., Sargent, W. L. W., & Peng, C. Y. 1997a, ApJS, 112, 391
Ho, L. C. & Peng, C. Y. 2001, ApJ, 555, 650
Ho, L. C., Rudnick, G., Rix, H., Shields, J. C., McIntosh, D. H., Filippenko, A. V., Sargent, W. L. W., & Eracleous, M. 2000, ApJ, 541, 120
Jarvis, M. J. & McLure, R. J. 2002, MNRAS, 336, L38
Kaspi, S., Smith, P. S., Netzer, H., Maoz, D., Jannuzi, B. T., & Giveon, U. 2000, ApJ, 533, 631
Kraemer, S. B., Ho, L. C., Crenshaw, D. M., Shields, J. C., & Filippenko, A. V. 1999, ApJ, 520, 564
Lacy, M., Laurent-Muehleisen, S. A., Ridgway, S. E., Becker, R. H., & White, R. L. 2001, ApJ, 551, L17
Laor, A. 1998, ApJ, 505, L83
– 2000, New Astronomy Review, 44, 503
Laor, A. & Draine, B. T. 1993, ApJ, 402, 441
Lara, L., Márquez, I., Cotton, W. D., Feretti, L., Giovannini, G., Marcaide, J. M., & Venturi, T. 1999, NewAR., 43, 643
Macchetto, F., Marconi, A., Axon, D. J., Capetti, A., Sparks, W., & Crane, P. 1997, ApJ, 489, 579
Maiolino, R., Salvati, M., Bassani, L., Dadina, M., della Ceca, R., Matt, G., Risaliti, G., & Zamorani, G. 1998, A&A, 338, 781
Matt, G., Fabian, A. C., Guainazzi, M., Iwasawa, K., Bassani, L., & Malaguti, G. 2000, MNRAS, 318, 173
McElroy, D. B. 1995, ApJS, 100, 105
McIntosh, D. H., Rieke, M. J., Rix, H.-W., Foltz, C. B., & Weymann, R. J. 1999, ApJ, 514, 40
Moran, E. C., Filippenko, A. V., Ho, L. C., Shields, J. C., Belloni, T., Comastri, A., Snowden, S. L., & Sramek, R. A. 1999, PASP, 111, 801
Moran, E. C., Barth, A. J., Kay, L. E., & Filippenko, A. V. 2000, ApJ, 540, L73
Moran, E. C., Kay, L. E., Davis, M., Filippenko, A. V., & Barth, A. J. 2001, ApJ, 556, L75
Murray, N., Chiang, J., Grossman, S. A., & Voit, G. M. 1995, ApJ, 451, 498
Nelson, C. H. & Whittle, M. 1995, ApJS, 99, 67
– 1996, ApJ, 465, 96
Nelson, C. H. 2000, ApJ, 544, L91
Netzer, H. & Laor, A. 1993, ApJ, 404, L51
Nicastro, F. 2000, ApJ, 530, L65
Oshlack, A. Y. K. N., Webster, R. L., & Whiting, M. T. 2002, ApJ, 576, 81
Panessa, F. & Bassani, L. 2002, A&A, 394, 435
Peterson, B. M. 1997, An introduction to active galactic nuclei (Cambridge, CUP)
Peterson, B. M. & Wandel, A. 2000, ApJ, 540, L13
Phinney, E. S. 1989, NATO ASIC Proc. 290: Theory of Accretion Disks, 457
Prugniel, Ph. & Simien, F. 1996, A&A, 309, 749
Rees, M. J., Phinney, E. S., Begelman, M. C., & Blandford, R. D. 1982, Nature, 295, 17
Rees, M. J., Netzer, H., & Ferland, G. J. 1989, ApJ, 347, 640
Risaliti, G., Maiolino, R., & Salvati, M. 1999, ApJ, 522, 157
Sabra, B. M., Shields, J. C., Ho, L. C., Barth, A. J., & Filippenko, A. V. 2003, ApJ, 584, 164
Sarzi, M., Rix, H., Shields, J. C., Rudnick, G., Ho, L. C., McIntosh, D. H., Filippenko, A. V., & Sargent, W. L. W. 2001, ApJ, 550, 65
Scoville, N. 2002, Active Galactic Nuclei: from Central Engine to Host Galaxy, Eds. S. Collin, F. Combes and I. Shlosman. (SF, ASP), 57
Singal, A. K. 1993, MNRAS, 262, L27
Shakura, N. I. & Sunyaev, R. A. 1973, A&A, 24, 337
Shields, J. C., Rix, H., McIntosh, D. H., Ho, L. C., Rudnick, G., Filippenko, A. V., Sargent, W. L. W., & Sarzi, M. 2000, ApJ, 534, L27
Shields, G. A., Gebhardt, K., Salviander, S., Wills, B. J., Xie, Brotherton, M. S., Yuan, J., & Dietrich, M. 2003, ApJ, 583, 124
Shlosman, I., Vitello, P. A., & Shaviv, G. 1985, ApJ, 294, 96
Smith, J. E., Young, S., Robinson, A., Corbett, E. A., Giannuzzo, M. E., Axon, D. J., & Hough, J. H. 2002, MNRAS, 335, 773
Tran, H. D. 1995, ApJ, 440, 597
– 2001, ApJ, 554, L19
– 2003, ApJ, 583, 632
Tremaine, S. et al. 2002, ApJ, 574, 740
Veilleux, S. 1991, ApJ, 369, 331
Veilleux, S., Sanders, D. B., & Kim, D.-C. 1997, ApJ, 484, 92
– 1999, ApJ, 522, 139
Wanders, I. & Peterson, B. M. 1996, ApJ, 466, 174
Whittle, M. 1985, MNRAS, 216, 817
Whysong, D. & Antonucci, R. R. J. 2003, ApJ, submitted (astro-ph/0207385).
Woo, J. & Urry, C. M. 2002, ApJ, 579, 530

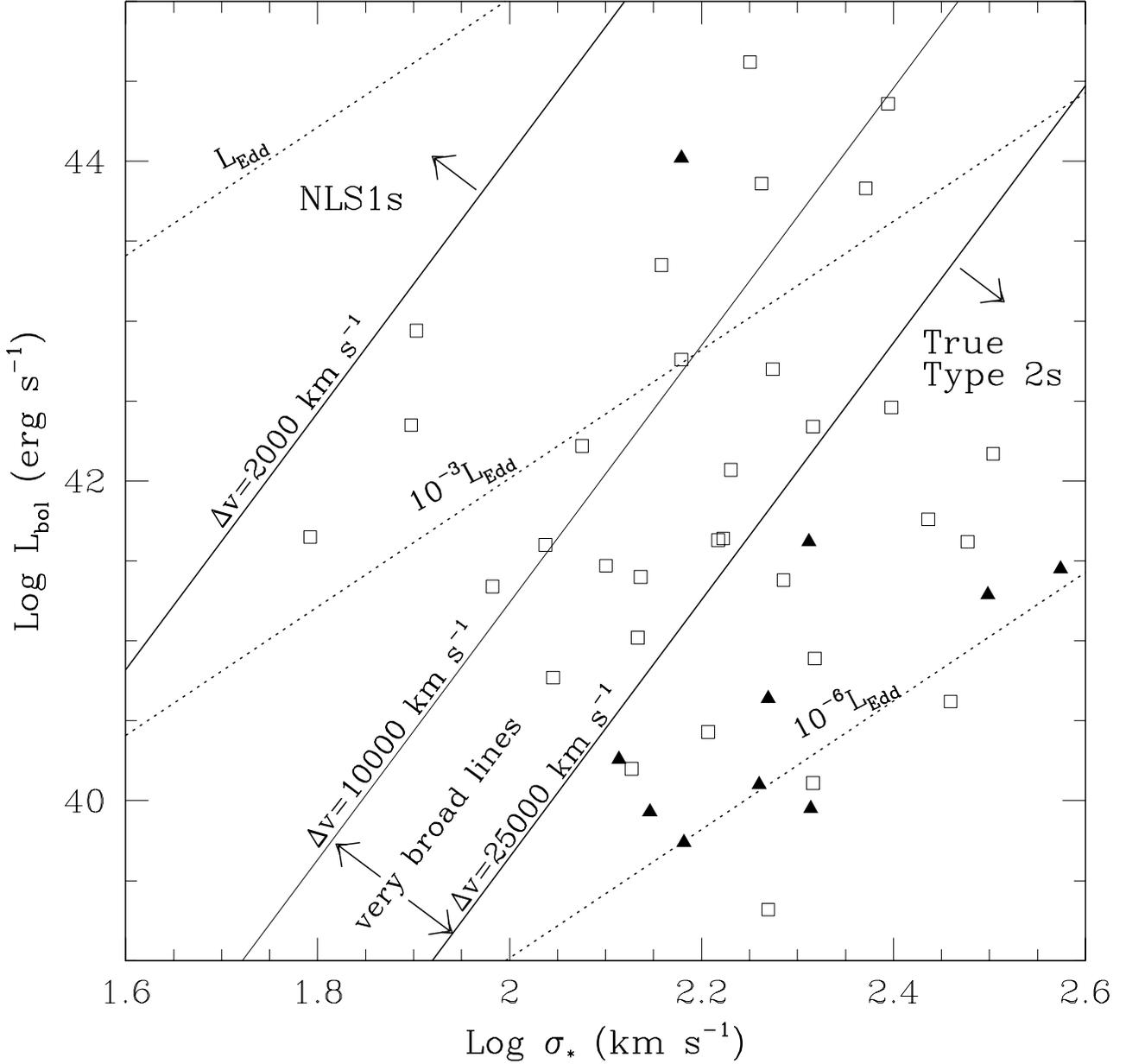

FIG. 1.— The distribution of the objects in Tables 1 (open squares) & 2 (filled triangles) in the $L_{bol}$ vs. $\sigma_*$ plane. The solid and dashed diagonal lines mark the position of objects with a given $\Delta v_{pred}$, or $\dot{m}$, as labeled. Objects with $\Delta v_{pred} > 25,000$ km s$^{-1}$ are likely to have no BLR emission, i.e. be true type 2 AGN. Objects with $\Delta v_{pred} = 10,000 - 25,000$ km s$^{-1}$ are likely to show extremely broad Balmer lines, which are often double peaked. The majority of luminous AGN ($L_{bol} > 10^{45}$ erg s$^{-1}$) fall in the range $\Delta v_{pred} = 2000 - 10,000$ km s$^{-1}$. AGN with $\Delta v_{obs} < 2000$ km s$^{-1}$ are commonly termd NLS1s. Luminous NLS1s must have $\dot{m} \lesssim 1$, but very low luminosity NLS1s can have $\dot{m} \ll 1$.

TABLE 1

Broad Hα FWHM for the Ho et al. Sample

| Galaxy | Class | $\sigma_*$ | $L_{bol}$ | $\dot{m}$ | $\Delta v_{pred}$ | $\Delta v_{obs}$ | Ref. |
|---|---|---|---|---|---|---|---|
| (1) | (2) | (3) | (4) | (5) | (6) | (7) | (8) |
| 315 | L1.9 | 300 | 41.62 | −5.32 | 73 | 2.0[a] | 1 |
| 1052 | L1.9 | 207 | 41.49 | −4.80 | 37 | 1.95[b] | 2 |
| 1161 | T1.9 | 288 | 40.62 | −6.24 | 120 | 3 | 1 |
| 1275 | S1.5 | 248 | 44.36 | −2.24 | 10 | 2.75[c] | 2 |
| 2639 | S1.9 | 188 | 42.70 | −3.42 | 15 | 3.1 | 1 |
| 2681 | L1.9 | 111 | ∼40.77 | ∼ −4.43 | ∼16 | 1.55 | 1 |
| 2985 | T1.9 | 134 | 40.20 | −5.33 | 33 | 2.05 | 1 |
| 3031 | S1.5 | 167 | 41.64 | −4.27 | 22 | 2.65[d] | 2 |
| 3226 | L1.9 | 208 | 40.89 | −5.41 | 53 | 2 | 1 |
| 3227 | S1.5 | 144 | 43.35 | −2.30 | 6.2 | 2.95 | 2 |
| 3516 | S1.2 | 235,156 | 43.83 | −2.68,−1.96 | 12.5,5.5 | 3.85 | 1,3 |
| 3642 | L1.9 | 137 | 41.40 | −4.17 | 17 | 1.25 | 1 |
| 3982 | S1.9 | 62 | 41.65 | −2.53 | 3 | 2.15 | 2 |
| 3998 | L1.9 | 319 | 42.17 | −4.87 | 60 | 2.15 | 2 |
| 4036 | L1.9 | 193 | 41.38 | −4.78 | 35 | 1.85 | 1 |
| 4051 | S1.2 | 80 | 42.94 | −1.68 | 2.4 | 1 | 4 |
| 4138 | S1.9 | 161 | 40.43 | −5.42 | 42 | 1.65 | 1 |
| 4151 | S1.5 | 178,93 | 44.62 | −1.40,−0.27 | 4.6,1.2 | 3.25 | 1,4 |
| 4168 | S1.9 | 186 | 39.32 | −6.78 | 105 | 2.85 | 1 |
| 4203 | L1.9 | 165,123 | 41.63 | −4.26,−3.75 | 22,12 | 1.5[e] | 1,5 |
| 4278 | L1.9 | 250 | 42.46 | −4.16 | 31 | 1.95 | 1 |
| 4388 | S1.9 | 119 | 42.22 | −3.10 | 8.1 | 3.9 | 1 |
| 4450 | L1.9 | 126 | 41.47 | −3.95 | 14 | 2.3[f] | 1 |
| 4565 | S1.9 | 136 | 41.02 | −4.53 | 21 | 2.3 | 1 |
| 4579 | S1.9 | 170 | 42.07 | −3.87 | 18 | 2.3[g] | 2 |
| 4636 | L1.9 | 207 | 40.11 | −6.18 | 83 | 2.45 | 1 |
| 4639 | S1.0 | 96 | 41.34 | −3.60 | 8.7 | 3.6 | 6 |
| 5033 | S1.5 | 151 | 42.76 | −2.98 | 9.5 | 2.85 | 6 |
| 5077 | L1.9 | 273 | 41.76 | −5.01 | 56 | 2.3 | 1 |
| 5273 | S1.5 | 79,52 | 42.35 | −2.26,−1.53 | 3.3,1.4 | 3.35 | 2,1 |
| 5548 | S1.5 | 183 | 43.86 | −2.21 | 7.5 | 4.2 | 1 |
| 7479 | S1.9 | 109 | 41.60 | −3.57 | 9.7 | 2.25 | 1 |

Notes.-Col.(1): Galaxy NGC number. Col.(2): Classification of nucleus from Ho et al. 1997b. Col.(3): Adopted $\sigma_*$ in km s$^{-1}$. Col.(4): Log $L_{bol}$ in units of erg s$^{-1}$, based on $M_B^{nuc}$ or $L(H\alpha)$. Col.(5): Log $\dot{m}$. Col.(6): Predicted Hα FWHM in 1000 km s$^{-1}$. Col.(7): Observed Hα FWHM in 1000 km s$^{-1}$, taken from Ho et al. 1997a. Col.(8): References for $\sigma_*$.
[a] Spectropolarimetry by Barth et al. (1999) gives $\Delta v \sim 3.2$.
[b] Spectropolarimetry by Barth et al. (1999) gives $\Delta v \sim 5$.
[c] Ho et al. (1997a) also find a very broad base component with full width near zero intensity of $\sim 19,000$ km s$^{-1}$.
[d] HST spectroscopy by Bower et al. (1996) shows a very broad base with FWZI of 11,500 km s$^{-1}$.
[e] HST spectroscopy by Shields et al. (2000) implies $\Delta v \sim 8.5$..
[f] HST spectroscopy by Ho et al. (2000) gives $\Delta v \sim 9.5$.
[g] HST spectroscopy by Barth et al. (2001a) implies $\Delta v \sim 8$.
References.- (1) McElroy 1995; (2) Nelson & Whittle 1995; (3) Prugniel & Simien (1996); (4) Ferrarese et al. 2001; (5) Sarzi et al. 2001; (6) Barth, Ho, & Sargent 2002.

TABLE 2

Broad Hα FWHM for the Tremaine et al. Sample

| Galaxy | Class | $\sigma_*$ | $L_{bol}$ | $M_{BH}$ | $\dot{m}$ | $\Delta v_{pred}$ | $\Delta v_{obs}$ |
|---|---|---|---|---|---|---|---|
| (1) | (2) | (3) | (4) | (5) | (6) | (7) | (8) |
| 1068 | S1.8 | 151 | 44.02 | 7.18 | −1.25 | 2.7 | 3.21 |
| 2787 | L1.9 | 140 | 39.93 | 7.61 | −5.78 | 47 | 2.5 |
| 3245 | T2: | 205 | 41.62 | 8.32 | −4.80 | 40 | ... |
| 3379 | L2/T2: | 206 | 39.95 | 8.00 | −6.15 | 73 | ... |
| 3608 | L2/S2: | 182 | 40.10 | 8.28 | −6.28 | 92 | ... |
| 4258 | S1.9 | 130 | 40.26 | 7.59 | −5.43 | 38 | 1.7 |
| 4261 | L2 | 315 | 41.29 | 8.72 | −5.53 | 77 | 2.8[a] |
| 4459 | T2: | 186 | 40.64 | 7.85 | −5.31 | 41 | ... |
| 4486 | L2 | 375 | 41.45 | 9.48 | −6.13 | 169 | 1.2[b] |
| 4596 | L2: | 152 | 39.74 | 7.89 | −6.25 | 72 | ... |

Notes.-Cols.(1-3): As in Table 1. Col.(4) As in Table 1, corrected for the distance in Tremaine et al. Col.(5): Log $M_{BH}$ in units of $M_\odot$, from Tremaine et al. Col.(6-8): As in Table 1.
[a] Measured in polarized light by Barth et al. (1999).
[b] From HST spectroscopy by Harms et al. (1994).